%% file: paper.tex
\documentclass[sigconf,nonacm,10pt]{acmart}
\usepackage[english]{babel}
\usepackage{blindtext}
\usepackage{wrapfig}
\usepackage{caption}
\usepackage{graphicx} 
\usepackage{float}    
\usepackage{stfloats} 
\usepackage{subcaption}

\usepackage[capitalize]{cleveref}
\usepackage{siunitx}

\acmDOI{}

\acmISBN{}

\author{Yang Qinyuan}
\affiliation{\country{Nanjing University of Aeronautics and Astronautics}}
\author{Hua Xiaoxiang}
\affiliation{\country{Nanjing University of Aeronautics and Astronautics}}
\author{Zhang Tong*}
\affiliation{\country{Nanjing University of Aeronautics and Astronautics}}
\author{Ren Fengyuan}
\affiliation{\country{Tsinghua University}}

\renewcommand{\shortauthors}{}

\begin{document}

\title{RDMACell: Token-Based Flowcell-Level RDMA Load Balancing for Large-Scale AI Training}

\begin{abstract}
Remote Direct Memory Access (RDMA) is a core technology for high-performance data center networks. However, its default Equal-Cost Multi-Path (ECMP) load-balancing mechanism often suffers from severe performance degradation due to hash collisions and elephant flows. Existing solutions have obvious limitations: flowlet-based approaches lack the necessary time gaps to trigger flowlet switching, while packet-level approaches suffer from severe packet reordering problems. On the other hand, although in-network programmable solutions can achieve fine-grained control, they are difficult to deploy in commercial general-purpose infrastructures.

This paper presents RDMACell, a host-side flowcell-level load-balancing system. It proactively splits flows into multiple flowcells, distributes them across multiple paths, and achieves real-time feedback through atomic dual Work Queue Elements (WQEs). With token-based receiver-side control, RDMACell enables microsecond-level path switching and extremely low packet loss without modifying NIC firmware or switch hardware. The ns-3 simulations in a fat-tree topology show that under 80\% network load with all-to-all traffic, RDMACell reduces 99th percentile FCT by 44\% compared with ECMP and by 42.2\% compared with LetFlow, while achieving tail latency performance comparable to state-of-the-art in-network solutions.
\end{abstract}

\maketitle

\input{body}

\bibliographystyle{ACM-Reference-Format}

\bibliography{reference}

\end{document}

%% file: body.tex
\section{Introduction}
In recent years, the explosive growth of Large Language Models (LLMs) has reshaped the fundamental architecture of data centers at an unprecedented pace, while imposing severe trans mission pressure on the underlying network~\cite{gangidi2024rdma,cao2024crux,hu2024characterization}. Modern LLMs represented by GPT-4 possess hundreds of billions or even trillions of parameters~\cite{hu2024characterization,jiang2024megascale,li2025revisiting}. Such large-scale models require parallel training on large computing clusters consisting of hundreds to thousands of Graphics Processing Units (GPUs) or Tensor Processing Units (TPUs)~\cite{xue2019fast,kalia2014using}.

In this distributed training paradigm, frequent gradient synchronization and parameter exchange produce intensive collective communications, including AllReduce, AllGather, and All‑to‑All~\cite{qian2024alibaba}. LLM training is highly sensitive to network bandwidth and tail latency; even small transmission delays can cause costly GPU idling. Thus, ultra‑high bandwidth, ultra‑low latency, and stable transmission have become essential requirements. Traffic patterns like All‑to‑All further aggravate network congestion and transmission pressure~\cite{cao2024crux,li2025revisiting}.

Currently, Remote Direct Memory Access (RDMA) has become the key technology for high‑performance data center networks, with advantages such as kernel bypass, zero‑copy, and hardware offloading~\cite{guo2016rdma}. To avoid expensive dedicated InfiniBand switches, academia and industry widely use RDMA over Converged Ethernet v2 (RoCEv2), which supports RDMA over standard Ethernet and reduces deployment costs~\cite{gao2021cloud}. However, in large‑scale deployments, RDMA still faces critical challenges in congestion control and multipath load balancing, limiting its performance in LLM training.

In practical deployment, RDMA adopts a connection-bound single-path transmission model and relies on Equal-Cost Multi-Path (ECMP) for path distribution ~\cite{hopps2000analysis,lapukhov2016use}. However, ECMP uses a static hashing mechanism, which easily leads to hash polarization for LLM training traffic characterized by bursty elephant flows~\cite{qian2024alibaba}. More importantly, RDMA traffic is characterized by small flow intervals, making it unsuitable for flowlet-based load balancing techniques~\cite{vanini2017let,katta2016clove,al2010hedera}. Existing per-packet load balancing schemes~\cite{ghorbani2017drill,dixit2013impact}  frequently cause severe packet reordering at the receiver in RoCEv2 environments. However, RoCEv2 strongly depends on lossless and strict in-order packet delivery. Once reordering occurs, the receiver RDMA Network Interface Controller (RNIC) immediately misinterprets it as packet loss and triggers retransmission. However, existing retransmission schemes, including Go-Back-N and Selective Retransmission, cannot effectively handle reordering and packet loss simultaneously: they either incur excessive redundant retransmissions or consume valuable on-chip memory on RNIC~\cite{mittal2018revisiting,wang2023srnic}. Even a small number of reordered packets can trigger heavy retransmissions, resulting in a sharp drop in network throughput and severe degradation of Flow Completion Time (FCT) performance.

In this paper, we present RDMACell, a novel token-based flowcell-level host-side load balancing system tailored for large-scale AI training. RDMACell elegantly offloads complex load-balancing logic to the host-side application layer for collaborative processing between the sender and receiver. Specifically, RDMACell is composed of several core mechanisms tailored for RDMA characteristics:

(1)Flowcell transmission model, which breaks away from traditional flow-level or aggressive packet-level load balancing models by proactively dividing data flows into fixed-size flowcells. It's typically set at 1.5 times the Bandwidth-Delay Product (BDP), which avoids the severe out-of-order packet issues inherent to RoCEv2 while preserving the fine granularity essential for dynamic load balancing.

(2) An atomic dual-WQE chain, a signaling WQE that uses immediate data to explicitly trigger a completion interrupt at the receiver to detect flowcell boundaries, while a data WQE silently transmits the payload to maximally preserve low CPU overhead.

(3) Token-based feedback. Upon detecting a signaling interrupt, the receiver generates a compact Token containing a global identifier and timestamp, writing it directly into a preallocated Token Slot in the sender's memory via a one-sided RDMA WRITE. By asynchronously polling these slots, the sender's scheduler achieves microsecond-scale path monitoring and dynamic timeout calculations without relying on programmable switches.

(4) Pipelined transmission and zero-copy congested path switching. While waiting for token feedback, RDMACell selects appropriate flows based on the current state to advance the transmission sliding window and maintains continuous data transmission, thus stably sustaining line-rate throughput. Meanwhile, the system is equipped with a congestion signal feedback mechanism that can rapidly detect congestion anomalies in the network. Once congestion is detected on a transmission path, the problematic path will be immediately isolated, and the system will automatically switch to the fast recovery mode, minimizing the impact of congestion on overall transmission.

While remaining fully compatible with the existing standard RoCEv2 protocol and commodity hardware, RDMACell achieves significant performance gains across diverse traffic patterns. In our ns-3 simulations involving a 128-node Fat-Tree topology, RDMACell reduces the 99th percentile FCT by up to 44\% compared to ECMP and 42.2\% compared to LetFlow with all-to-all traffic, while achieving tail latency performance comparable to state-of-the-art in-network solutions. 

\begin{figure}[!htbp]  
  \centering  
  \includegraphics[width=0.45\textwidth]{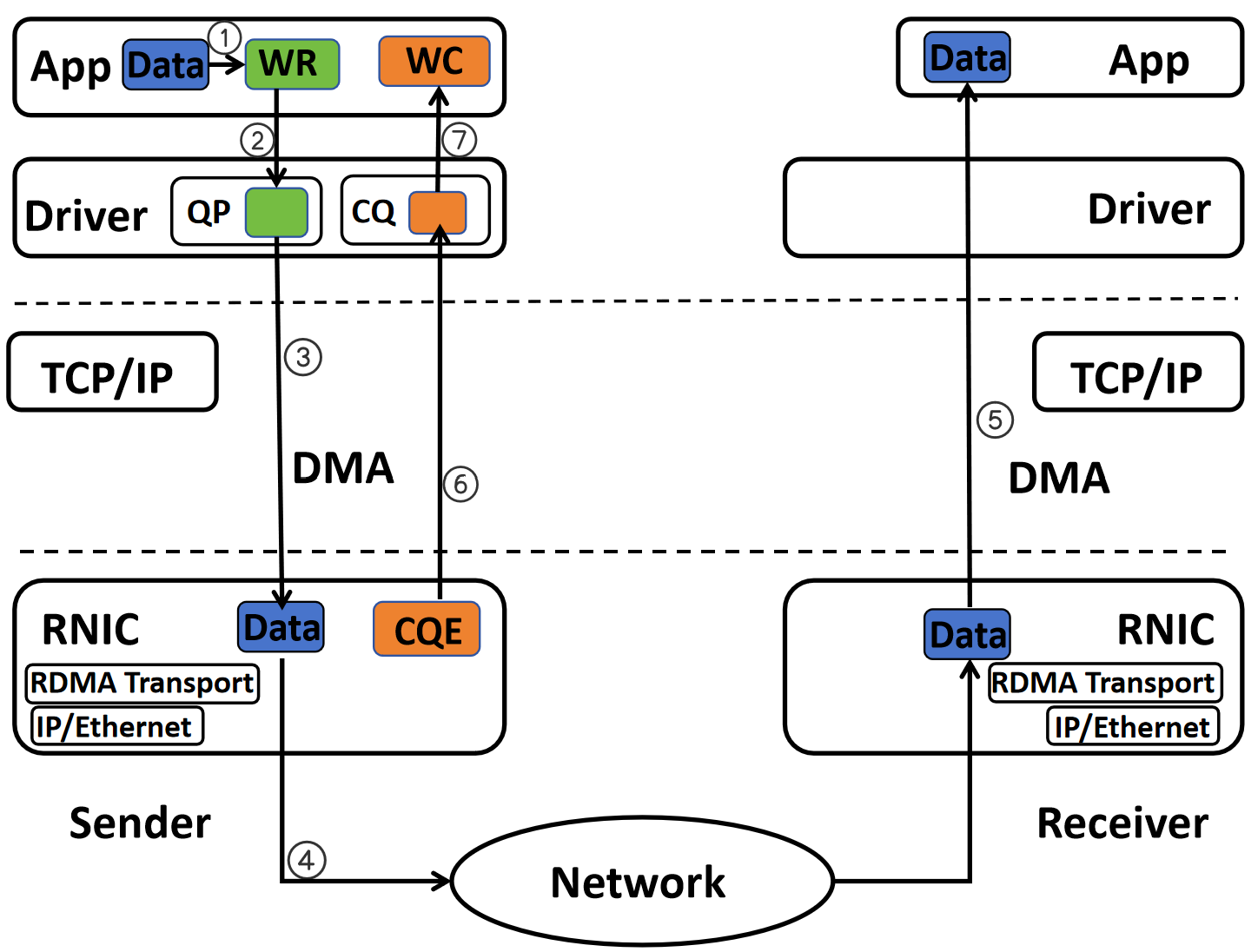} 
  \caption{The workflow of RDMA transport} 
  \label{fig:rdma_workflow} 
\end{figure}
\section{Background and Motivation}
\subsection{RDMA Overview} 
The core communication abstraction of RDMA is the Queue Pair (QP), which consists of a Send Queue (SQ) and a Receive Queue (RQ). Applications initiate data transfers by submitting Work Queue Elements (WQEs), bypassing the operating system (OS).

As illustrated in Fig.\ref{fig:rdma_workflow}, the sender creates a Work Request (WR) for each message to be transmitted, registering relevant message metadata within the WR. This request is subsequently posted to the SQ via the ibv\_post\_send() API. The RNIC processes WRs in the SQ sequentially, accessing memory buffers through DMA operations. Upon completion, the RNIC optionally pushes a Completion Queue Element (CQE) to the associated Completion Queue (CQ), allowing the application to perceive the operation's status.

Modern RDMA deployments predominantly rely on the RoCEv2 protocol, which encapsulates RDMA packets within UDP/IP headers. This design enables seamless transmission over standard Ethernet. RDMA supports three primary transport modes: Reliable Connected (RC), Unreliable Connected (UC), and Unreliable Datagram (UD). Although both RC and UC require connection establishment between hosts, RC guarantees in-order and reliable delivery, whereas UC does not account for packet loss. UD operates in a connectionless manner and provides no guarantees for ordering or delivery. Currently, RC is the dominant transport mode in bulk data transmission due to its hardware-offloaded reliable in-order delivery~\cite{gangidi2024rdma}, so this study focuses on the RC mode. However, the RC mode is highly sensitive to out-of-order packets, which trigger inefficient hardware-level Go-Back-N retransmissions. Furthermore, unlike TCP, commercial RNICs typically implement hardware-level packet pacing. This underlying mechanism ensures that RDMA traffic exhibits high degrees of smoothness and continuity.

\subsection{Limitations of Existing Load Balancing}
ECMP remains widely deployed in modern data centers. By performing a static hash on the flow five-tuple, ECMP maps flows to equivalent egress links. However, in modern applications such as AI training, the prevalence of elephant flows frequently leads to hash polarization~\cite{qian2024alibaba}. Multiple elephant flows may be assigned to the same path, resulting in severe local congestion and Head-of-Line (HOL) blocking, while other network links remain underutilized.
To address these challenges, academia and industry have proposed optimization schemes. Flowlet-based schemes like CONGA~\cite{alizadeh2014conga} let switches detect congestion and pick optimal paths per flowlet. However, gathering congestion feedback takes time, and rerouting depends on idle intervals. LetFlow~\cite{vanini2017let} performs well in TCP by using natural traffic gaps, but switching links takes two RTTs. In RDMA networks, RNIC enables smooth packet transmission, so there are no idle intervals for passive flowlet switching. Thus, such schemes resemble ECMP.

Other existing solutions attempt to implement effective load balancing in general commercial RDMA environments: MP-RDMA~\cite{lu2018multi} achieves multipath transmission through multipath ACK clocking, reorder-aware path selection, and synchronization mechanisms, but requires modifications to the underlying logic of existing RNICs. ConWeave~\cite{song2023network} reduces reordering issues and achieves path rerouting through queue control on programmable switches. Unfortunately, it suffers from insufficient flexibility under high-load scenarios due to constrained rerouting conditions. Under collective communication in large-scale model training, these in-network solutions introduce high deployment costs. In summary, existing load-balancing technologies either fail to trigger flexible path switching or lack generalizability due to a heavy reliance on specialized programmable switches.

Therefore, there is an urgent need for a purely host-based, fine-grained load-balancing architecture. This architecture should proactively break the constraint of RDMA traffic smoothness without modifying commodity RNIC firmware or switch hardware. It should also achieve microsecond-level congestion awareness, flexible multi-path switching, and zero-copy fault recovery with minimal system overhead. These requirements form the core motivation for the proposed RDMACell in this paper.

\section{RDMACell Design}

\begin{figure}[htbp]  
  \centering  
  \includegraphics[width=0.45\textwidth]{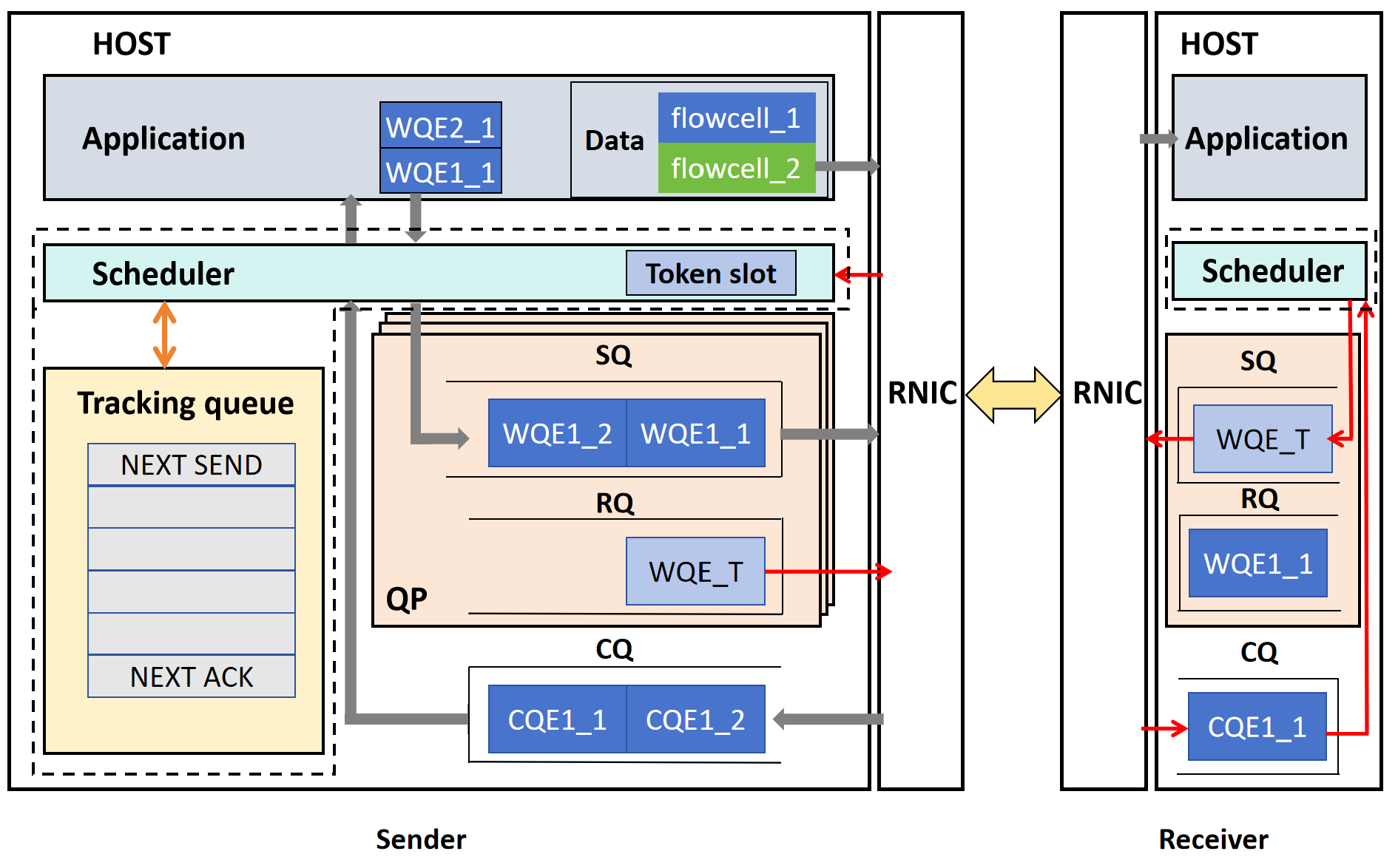} 
  \caption{The architecture of RDMACell} 
  \label{fig:RDMAcell_arch} 
\end{figure}
As shown in Fig.\ref{fig:RDMAcell_arch}, RDMACell implements a user-space transmission layer consisting of the scheduler and the tracking queue. The scheduler serves as the execution engine, achieving real-time path awareness and $T_{soft}$ fast recovery by polling token slots. Meanwhile, the tracking queue maintains a sliding window via NEXT SEND/ACK pointers, responsible for flowcell state tracking and providing the essential metadata for zero-copy retransmission. This section elaborates on its system architecture and key mechanisms: we first present the flowcell-based transmission protocol built on atomic dual-WQE chains and Token feedback (Section~\ref{sec:1}); then dissect the adaptive scheduling state machine that drives the system (Section~\ref{sec:2}); finally, we demonstrate the system’s compatibility and deployability on commodity RNICs without any hardware modifications (Section~\ref{sec:3}).

\subsection{Fine-grained Transmission Protocol}
\label{sec:1}
The RDMA protocol stack offers rich communication semantics, which are primarily divided into the SEND/ RECV (Two-sided) model, and the WRITE/ READ (One-sided) model for unilateral operations. Although SEND/RECV semantics are intuitive for handling small control packets, they require the receiver to pre-consume corresponding Receive Queue Elements (RQ WQEs) before processing. In high-concurrency, large-scale AI training scenarios, this easily leads to excessive CPU interrupt overhead and receive queue exhaustion. In contrast, one-sided operations bypass the remote CPU for higher throughput and lower latency. In large-scale AI training, the data communication phase is dominated by one-sided WRITE operations most of the time.  However, standard RDMA WRITE lacks Data Boundary Awareness; the receiver's hardware does not trigger a Completion Queue Event (CQE) upon arrival, preventing the receiver from detecting when a transfer is complete.

To address these issues, RDMACell abandons coarse-grained flow-level load balancing and aggressive packet-level load balancing and introduces flowcell~\cite{he2015presto} as the basic unit of scheduling and retransmission. Besides, its dual-WQE chain maintains the high performance of one-sided operations. The tracking queue achieves fine-grained tracking of flowcell states. The protocol includes the following core designs: By sizing the flowcell at $1.5 \times BDP$, we ensure the pipeline remains full while waiting for token feedback. This bounded size prevents massive data bursts from overflowing switch buffers and triggering PFC pause frames. To enable precise tracking of transmission units without violating standard standard RDMA hardware offloading, RDMACell completely avoids modifying the hardware-generated RDMA Base Transport Header (BTH). Instead, it embeds the essential in-band metadata, specifically the globally unique identifier (), directly into the 32-bit Immediate Data field of the accompanying signaling WQE. This enables precise tracking of transmission units without introducing any additional bandwidth overhead Global\_Cell\_ID.

\begin{figure}[htbp]  
  \centering  
  \includegraphics[width=0.45\textwidth]{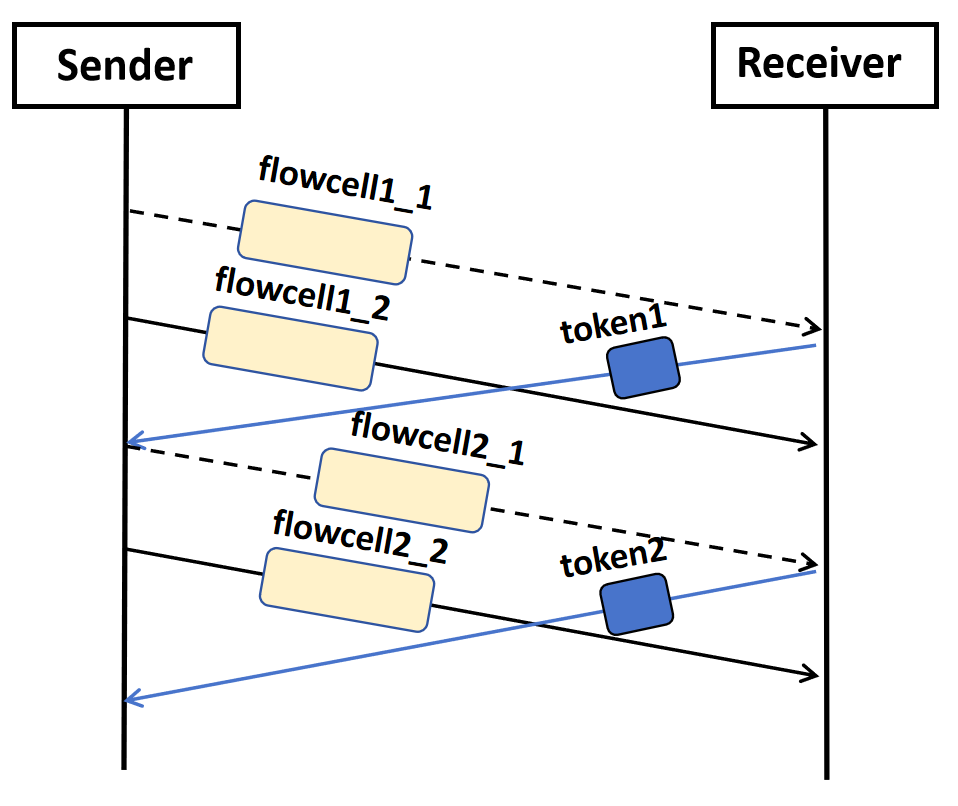} 
  \caption{RDMACell transmission process} 
  \label{fig:send} 
\end{figure}

To solve the lack of data boundary awareness in standard RDMA Write operations, as shown in Fig.\ref{fig:send}, RDMACell uses the Verbs-linked WQE mechanism to post an atomic dual-WQE chain: WQE-Token uses WRITE\_WITH\_IMM semantics and typically carries only 1 MTU of data. Its core role is to act as a signaling source. Crucially, the sender embeds the Global\_Cell\_ID entirely within the 32-bit immediate data of this operation. This explicitly triggers a completion interrupt at the receiver, allowing the receiver to precisely detect flowcell boundaries, extract the identifier directly from the completion queue element (CQE), and generate a Token. WQE-Payload follows WQE-Token and uses standard RDMA\_WRITE semantics to transmit the remaining payload of the flowcell. As a silent transmission, it does not impose additional CQE pressure on the receiver, thus preserving the low CPU overhead of RDMA one-sided operations. With this design, RDMACell converts each flowcell into an observable, schedulable transmission unit based on Token feedback. 

Upon receiving the immediate data interrupt from the signaling WQE, the receiver extracts the globally unique identifier of the flowcell and records the timestamp synchronously. It then generates a compact token containing the above information and issues a one-sided RDMA WRITE operation over the data QP. This token is written directly into a preallocated token slot region in the sender’s memory, which is logically organized as a fixed-size ring buffer, where each slot corresponds to the state of a specific flowcell. By utilizing the one-sided write operation of RDMA, the update process of tokens is automatically completed by the hardware DMA engine of the sending network card standard, and the data directly falls into the predefined memory area. 

To enable fine-grained path management, RDMACell deploys a tracking queue and a scheduler collaboratively at the sender side. The tracking queue is a data structure that maintains a sliding window. It precisely defines the state of data flows via its internal NEXT SEND and NEXT ACK pointers. The queue not only indicates the next flowcell to be sent, but also tracks all flowcells that have been transmitted yet not acknowledged by the CQ.

Meanwhile, the scheduler drives the entire system in a decoupled asynchronous working mode: it continuously polls the token slots in the memory to obtain real-time feedback from the receiver. Upon receiving a token, the scheduler advances the sliding window of the tracking queue using the token's ID and accurately calculates the RTT of the path with the token's timestamp to dynamically adjust the $T_{\text{soft}}$ timeout threshold.
\begin{equation}
\label{eq:soft_t}  
T_{\text{soft}} = \text{RTT}_{\text{avg}} + 2 \times \text{RTT}_{\text{var}}
\end{equation}
Among them, $RTT_{avg}$ is the average round-trip time, and $RTT_{var}$ is the absolute average deviation between RTT samples and the average value.

\begin{equation}
\label{eq:rtt_var_update}  
\text{RTT}_{\text{var}} = (1 - \beta) \text{RTT}_{\text{var}} + \beta \left| \text{RTT}_{\text{sample}} - \text{RTT}_{\text{avg}} \right|
\end{equation}
where $\beta$ serves as a smoothing coefficient to make $RTT_{var}$ smoother and more stable, is empirically set to 1/4, $RTT_{sample}$ is the round-trip time for the current transaction.

\subsection{Adaptive Scheduling State Machine}
\label{sec:2}
The RDMACell scheduler employs an adaptive state machine that enables low-latency, non-blocking switching between normal transfer and fast recovery states. With microsecond-level path switching and zero-copy recovery mechanisms, it addresses slow failure recovery and high retransmission overhead in high-speed networks.

Normal transfer is the steady-state operating mode. After receiving a token within the dynamic timeout threshold $T_{\text{soft}}$, RDMACell maintains the primary QP, advances the pending pointer of the Work Request Queue, updates the average RTT baseline, and sustains line-rate throughput. As long as the path is normal, the state machine remains in the steady state to maximize throughput and transmission efficiency.

Fast recovery is activated by explicit Negative Acknowledgment (NACK) feedback or timeout. Upon detecting an anomaly, the system immediately isolates the faulty QP and rolls back the pending pointer to the earliest unacknowledged flowcell. Leveraging the design of the tracking queue, RDMACell uses "side-channel recovery" to batch-submit relevant descriptors to operational backup QPs, with zero data copying throughout the process. Meanwhile, the faulty QP is reset asynchronously to break the hardware Go-Back-N retransmission loops and is later reconstructed for reuse, ensuring high reliability without sacrificing performance.

\subsection{Zero-Hardware-Modification}
\label{sec:3}
A major advantage of RDMACell is its lightweight, pure user-space implementation, which leverages the standard RDMA Verbs API to ensure full compatibility with commodity RNICs (e.g., Mellanox ConnectX series) without requiring invasive firmware or protocol stack modifications. 

By creating QPs via the standard \texttt{ibv\_create\_qp} interface and utilizing the UDP source port field in RoCEv2, the system maintains seamless compatibility with ECMP static hashing on underlying network switches. 

Furthermore, RDMACell ensures semantic compatibility by posting dual-WQE chains through standard \texttt{ibv\_post\_send} operations, where both immediate data and payloads strictly adhere to RDMA WRITE semantics for native hardware parsing at the receiver. This architecture is based on a zero-copy memory management scheme: application buffers and receiver memory pools are pre-registered using \texttt{ibv\_reg\_mr}, while the tracking queue manipulates only pointers and address offsets to fully exploit RDMACell's kernel-bypassing DMA mechanism.

\section{Evaluation}
\subsection{Simulation setup}
In this section, we conduct a performance evaluation of RDMACell through large-scale ns-3 simulations.

\textbf{Simulation Platform}: We build a high-precision RoCEv2 protocol stack based on the ns-3 platform, and implement the user-space scheduling engine, tracing queue, and dual-WQE of RDMACell.

\begin{figure}[htbp]  
  \centering  
  \includegraphics[width=0.45\textwidth]{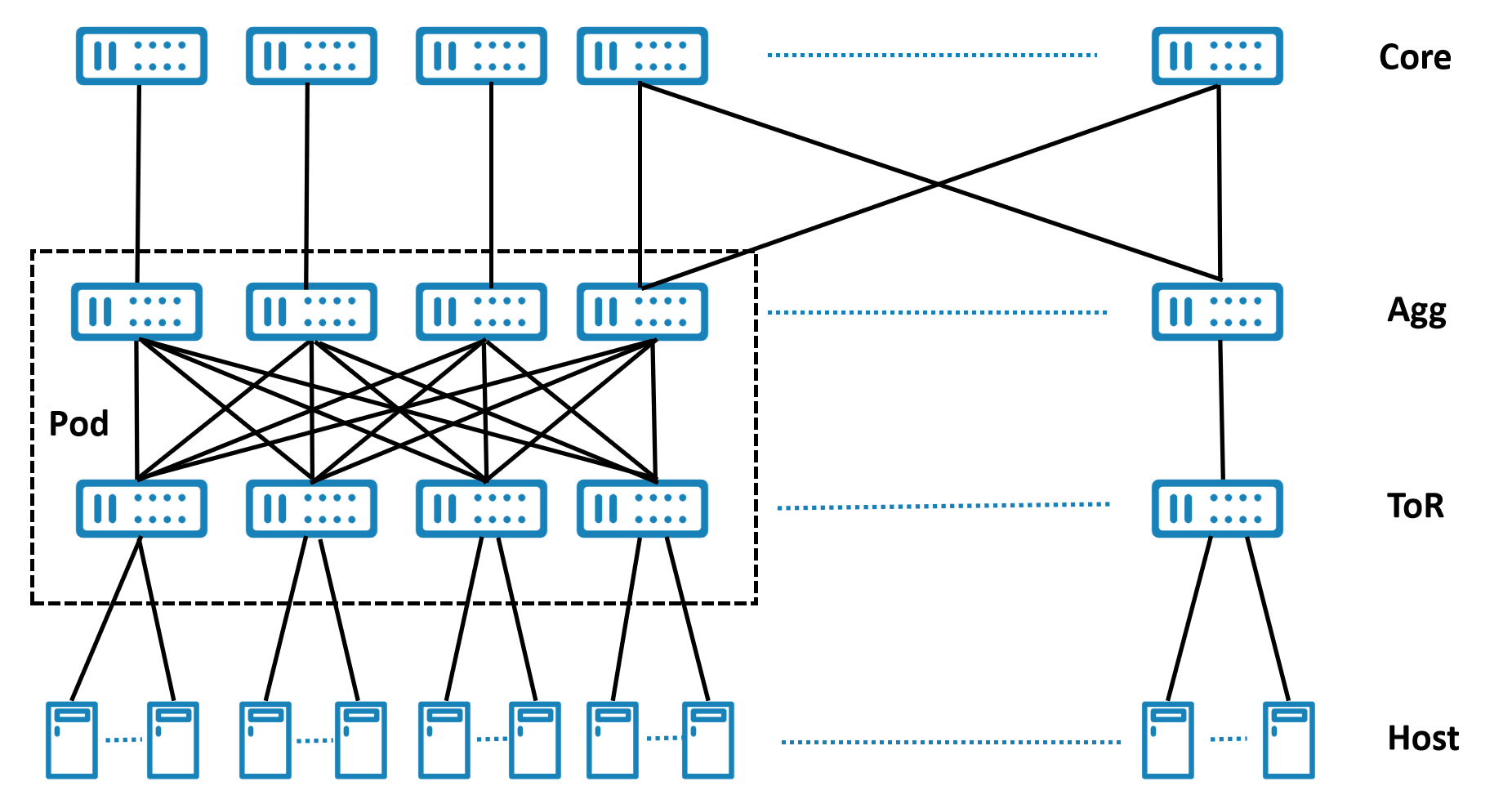} 
  \caption{Fat-tree topology used in simulations} 
  \label{fig:tp} 
\end{figure}

\textbf{Network Topology}: As shown in Fig.\ref{fig:tp}, a typical fat-tree topology with K=8 is adopted, consisting of 128 servers. All links in the core, aggregation, and access layers are uniformly configured to 100\,Gbps. The per-hop link latency is set to \SI{1}{\micro\second}. Switches employ a standard shared-buffered architecture with a per-port buffer size of 2\,MB, and the ECN marking function is enabled. The in-network state collection logic required by the compared algorithms (e.g., CONGA, ConWeave, HULA) has been implemented in the switch model following the original papers.

\textbf{Workloads}: We use real-world traffic traces from AliCloud Storage and Solar as workloads. 1) AliCloude Storage: Dominated by small flows, with a small number of medium-to-large flows forming a long tail, presenting a "small-flow dominated + long-tail" pattern~\cite{li2019hpcc}. 2) Solar: Highly small-flow oriented with an extremely short tail, presenting a "pure small-flow" pattern.

\textbf{Comparison Schemes}: We select mainstream load balancing schemes in modern data center networks for comparison, including ECMP, CONGA, LetFlow, HULA~\cite{katta2016hula}, and ConWeave.

\begin{figure*}[t]
  \centering
  
  \begin{subfigure}[b]{0.48\textwidth}
    \centering
    \includegraphics[width=\linewidth]{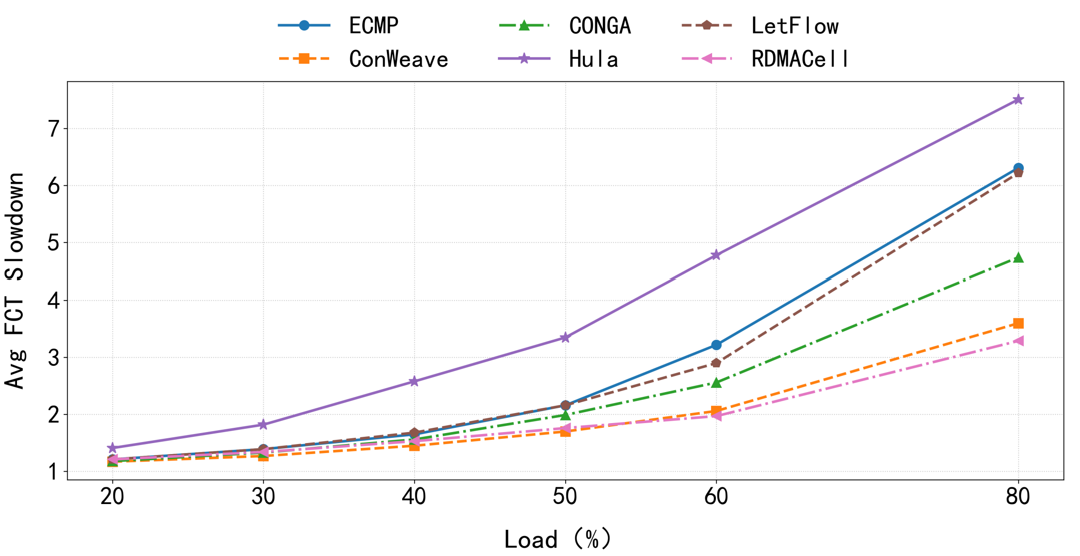}
    \caption{Avg FCT for AliCloud Storage workload}
    \label{fig:subfig1}
  \end{subfigure}
  \hfill 
  \begin{subfigure}[b]{0.48\textwidth}
    \centering
    \includegraphics[width=\linewidth]{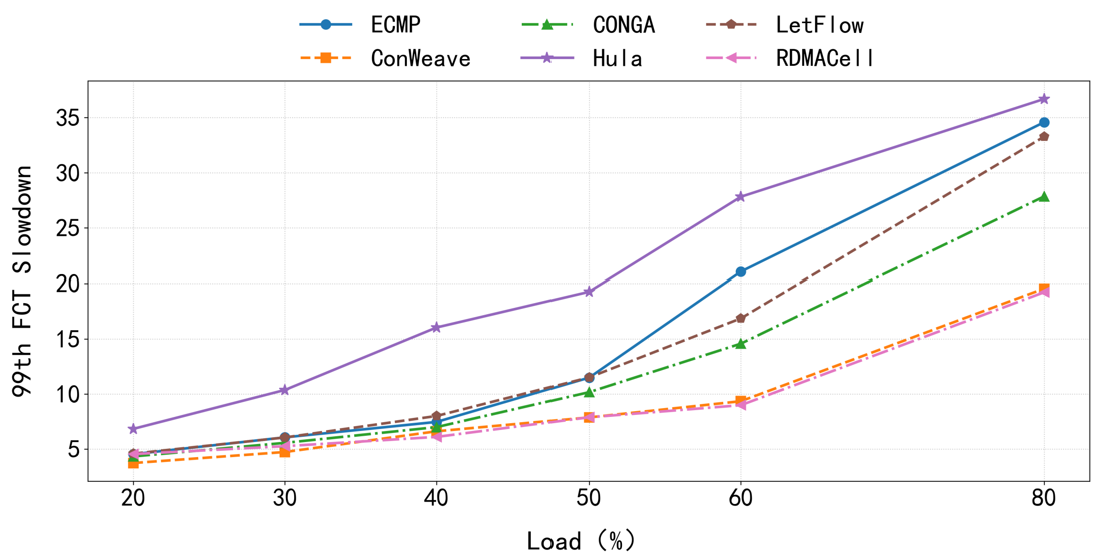}
    \caption{99th percentile FCT for AliCloud Storage workload}
    \label{fig:subfig2}
  \end{subfigure}

  \vspace{0.5cm} 

  \begin{subfigure}[b]{0.48\textwidth}
    \centering
    \includegraphics[width=\linewidth]{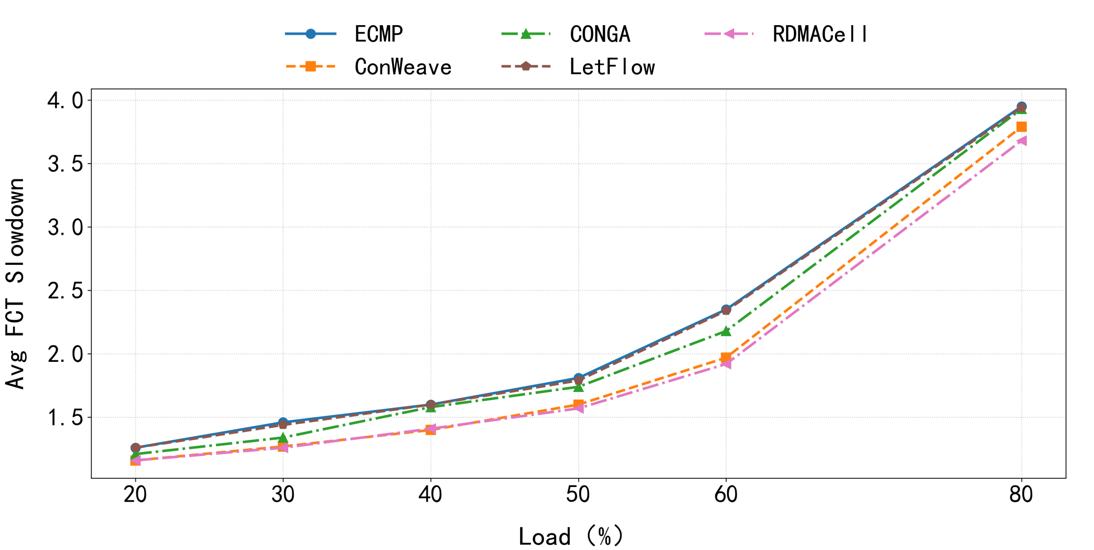}
    \caption{Avg FCT for Solar workload}
    \label{fig:subfig3}
  \end{subfigure}
  \hfill 
  \begin{subfigure}[b]{0.48\textwidth}
    \centering
    \includegraphics[width=\linewidth]{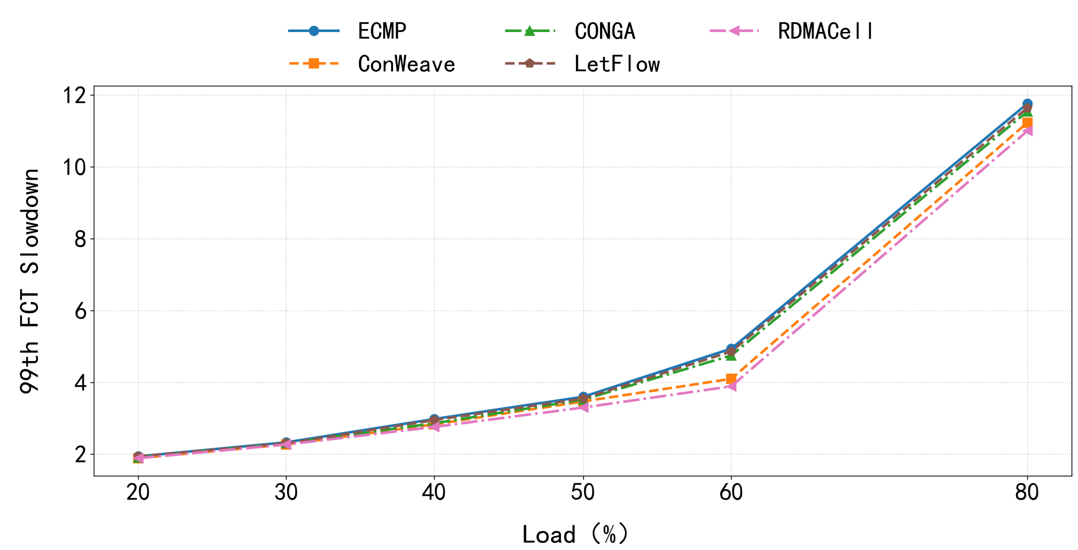}
    \caption{99th percentile FCT for Solar workload}
    \label{fig:subfig4}
  \end{subfigure}

  \caption{Average and 99th percentile FCT in all-to-all pattern.}
  \label{fig:mainfig}
\end{figure*}

\subsection{Results and Analysis}
This work presents an in-depth performance evaluation of RDMACell against five state-of-the-art load-balancing schemes: ECMP, ConWeave, CONGA, Hula, and LetFlow, under diverse traffic patterns, based on the ns-3 simulation platform. We emulate the typical data center All-to-all communication pattern to evaluate the adaptability of each scheme when handling Alibaba Cloud storage traffic (with long-tail characteristics) and Solar traffic (predominantly small flows). By gradually increasing the network load from 20\% to 80\%, we measure the average FCT and 99th percentile FCT to quantify overall efficiency and tail-latency stability.

Overall, under both traffic patterns, the Avg FCT and 99th percentile FCT of all schemes increase with network load, and traffic characteristics impose a significant impact on performance. As shown in Fig. \ref{fig:subfig1} and \ref{fig:subfig2}, in the Alibaba Cloud storage traffic scenario, the performance gap between schemes widens dramatically as load rises. At 80\% load, RDMACell achieves the best Avg FCT of 3.29, representing a 56.2\% reduction compared with HULA (7.51), which performs the worst. For the critical metric of 99th percentile FCT, RDMACell (19.23) reduces the tail latency by approximately 44\% compared with ECMP (34.58) and by 42.2\% compared with LetFlow (33.27), and by 47.1\% compared with Hula (36.69). Meanwhile, it outperforms ConWeave, a recent in-network reordering scheme based on programmable switches. The superior performance of RDMACell stems from its end-to-end congestion awareness and precise Flowcell rerouting strategy, which enable it to sensitively avoid path congestion caused by large flows and effectively mitigate the long-tail effect induced by queuing delay.
As shown in Fig. \ref{fig:subfig3} and \ref{fig:subfig4}, in the Solar traffic scenario, due to its short-tail nature, all schemes achieve lower and more convergent FCT values. At 80\% load, RDMACell reduces Avg FCT by only 6.8\% and 99th percentile FCT by 6.5\% compared with ECMP. Under light loads ($\leq$ 60\%), RDMACell demonstrates exceptional stability, with Avg FCT maintained between 1.16 and 1.88. Its 99th percentile FCT (3.93) is reduced by 20.4\% compared with ECMP (4.94). Even under extremely high load, its tail latency (11.00) remains competitive, verifying the scheme’s high efficiency in pure small-flow environments. Notably, HULA performs poorly under both traffic patterns. In Solar traffic, its performance falls far behind other schemes (Avg FCT exceeds 3 and 99th percentile FCT exceeds 15 at 30\% load, rising to over 30 under high load). Thus, its data is omitted from some result figures for clarity. The weak performance of HULA is mainly attributed to perception lag in its probe-based global state maintenance, which leads to outdated routing decisions that exacerbate link congestion when facing highly volatile and collision-prone All-to-all traffic.

In summary, RDMACell exhibits strong robustness under diverse traffic characteristics. These results validate its promising application potential in modern data center networks.

\section{Conclusion}
This paper proposes RDMACell, a pure host-side fine-grained load-balancing system for RDMA networks. RDMACell actively constructs and schedules flowcells via an atomic dual-WQE chain and a token feedback mechanism, enabling microsecond level path switching and congestion avoidance. Without modifying the firmware of commercial RNICs or switch hardware, RDMACell achieves high-performance load balancing using a purely software-defined scheduling strategy. Simulation results in a 128-node fat-tree topology show that RDMACell consistently minimizes tail latency and yields the minimal Avg FCT across diverse workloads, reducing the 99th percentile FCT by up to 44\% and 42.2\% compared to ECMP and LetFlow, while achieving tail latency performance comparable to state-of-the-art in-network solutions, providing a novel evolutionary direction for optimizing network load balancing in large-scale distributed AI training and high-performance storage environments.